\newcommand\apjcls{1}
\newcommand\aastexcls{2}
\newcommand\othercls{3}

\newcommand\papercls{\aastexcls}
\documentclass[tighten, times, preprint2]{aastex6}  

\if\papercls \apjcls
\usepackage{apjfonts}
\else\if\papercls \othercls
\usepackage{epsfig}
\fi\fi
\usepackage{ifthen}
\usepackage{natbib}
\usepackage{amssymb, amsmath}
\usepackage{appendix}
\usepackage{etoolbox}
\usepackage[T1]{fontenc}
\usepackage{paralist}

\if\papercls \apjcls
\newcommand\aas{\ref@jnl{AAS Meeting Abstracts}}
\newcommand\dps{\ref@jnl{AAS/DPS Meeting Abstracts}}
\newcommand\maps{\ref@jnl{MAPS}}
\else\if\papercls \othercls
\usepackage{astjnlabbrev-jh}
\fi\fi

\if\papercls \aastexcls
\hypersetup{citecolor=blue, 
            linkcolor=blue, 
            menucolor=blue, 
            urlcolor=blue}  
\else
\usepackage[
bookmarks=true,           
bookmarksnumbered=true,   
colorlinks=true,          
citecolor=blue,           
linkcolor=blue,           
menucolor=blue,           
urlcolor=blue,            
linkbordercolor={0 0 1},  
pdfborder={0 0 1},
frenchlinks=true]{hyperref}
\fi

\if\papercls \othercls
\newcommand{\eprint}[1]{\href{http://arxiv.org/abs/#1}{#1}}
\else
\renewcommand{\eprint}[1]{\href{http://arxiv.org/abs/#1}{#1}}
\fi

\providecommand{\adsurl}[1]{\href{#1}{ADS}}

\makeatletter
\patchcmd{\NAT@citex}
  {\@citea\NAT@hyper@{%
     \NAT@nmfmt{\NAT@nm}%
     \hyper@natlinkbreak{\NAT@aysep\NAT@spacechar}{\@citeb\@extra@b@citeb}%
     \NAT@date}}
  {\@citea\NAT@nmfmt{\NAT@nm}%
   \NAT@aysep\NAT@spacechar\NAT@hyper@{\NAT@date}}{}{}

\patchcmd{\NAT@citex}
  {\@citea\NAT@hyper@{%
     \NAT@nmfmt{\NAT@nm}%
     \hyper@natlinkbreak{\NAT@spacechar\NAT@@open\if*#1*\else#1\NAT@spacechar\fi}%
       {\@citeb\@extra@b@citeb}%
     \NAT@date}}
  {\@citea\NAT@nmfmt{\NAT@nm}%
   \NAT@spacechar\NAT@@open\if*#1*\else#1\NAT@spacechar\fi\NAT@hyper@{\NAT@date}}
  {}{}
\makeatother

\makeatletter
\DeclareRobustCommand{\lowcase}[1]{\@lowcase#1\@nil}
\def\@lowcase#1\@nil{\if\relax#1\relax\else\MakeLowercase{#1}\fi}
\makeatother

\DeclareSymbolFont{UPM}{U}{eur}{m}{n}
\DeclareMathSymbol{\umu}{0}{UPM}{"16}
\let\oldumu=\umu
\renewcommand\umu{\ifmmode\oldumu\else\math{\oldumu}\fi}

\if\papercls \othercls

\else

\fi

\let\oldsim=\sim
\renewcommand\sim{\ifmmode\oldsim\else\math{\oldsim}\fi}
\let\oldpm=\pm
\renewcommand\pm{\ifmmode\oldpm\else\math{\oldpm}\fi}
\newcommand\by{\ifmmode\times\else\math{\times}\fi}

\newcommand\tablebox[1]{\begin{tabular}[t]{@{}l@{}}#1\end{tabular}}
\newbox{\wdbox}
\renewcommand\c{\setbox\wdbox=\hbox{,}\hspace{\wd\wdbox}}
\renewcommand\i{\setbox\wdbox=\hbox{i}\hspace{\wd\wdbox}}

\newcount\timect
\newcount\hourct
\newcount\minct
\newcommand\now{\timect=\time \divide\timect by 60
         \hourct=\timect \multiply\hourct by 60
         \minct=\time \advance\minct by -\hourct
         \number\timect:\ifnum \minct < 10 0\fi\number\minct}

\catcode`@=11

\newcommand\comment[1]{}

\newcommand\commenton{\catcode`\%=14}

\renewcommand\math[1]{$#1$}
\newcommand\mathshifton{\catcode`\$=3}

\let\atab=&
\newcommand\atabon{\catcode`\&=4}

\let\oldmsp=\sp
\let\oldmsb=\sb
\def\sp#1{\ifmmode
           \oldmsp{#1}%
         \else\strut\raise.85ex\hbox{\scriptsize #1}\fi}
\def\sb#1{\ifmmode
           \oldmsb{#1}%
         \else\strut\raise-.54ex\hbox{\scriptsize #1}\fi}
\newbox\@sp
\newbox\@sb
\def\sbp#1#2{\ifmmode%
           \oldmsb{#1}\oldmsp{#2}%
         \else
           \setbox\@sb=\hbox{\sb{#1}}%
           \setbox\@sp=\hbox{\sp{#2}}%
           \rlap{\copy\@sb}\copy\@sp
           \ifdim \wd\@sb >\wd\@sp
             \hskip -\wd\@sp \hskip \wd\@sb
           \fi
        \fi}
\def\msp#1{\ifmmode
           \oldmsp{#1}
         \else \math{\oldmsp{#1}}\fi}
\def\msb#1{\ifmmode
           \oldmsb{#1}
         \else \math{\oldmsb{#1}}\fi}

\def\supon{\catcode`\^=7}

\def\subon{\catcode`\_=8}

\def\supsubon{\supon \subon}

\newcommand\actcharon{\catcode`\~=13}

\newcommand\paramon{\catcode`\#=6}

\comment{And now to turn us totally on and off...}

\newcommand\reservedcharson{ \commenton  \mathshifton  \atabon  \supsubon 
                             \actcharon  \paramon}

\catcode`@=12
\reservedcharson

\if\papercls \apjcls

\else

\fi

\if\papercls \othercls
\else
  \newcommand\inpress{n}
  \if\inpress y
    \received{\today}
    \revised{}
    \accepted{}
    \if\papercls \apjcls
    \slugcomment{}
    \fi
  \else
  \slugcomment{\tablebox{In preparation for {\em ApJ}. DRAFT of {\today}.}}
  \fi
\fi

\newcommand\chisq{\ifmmode{\chi\sp{2}}\else\math{\chi\sp{2}}\fi}
\newcommand\redchisq{\ifmmode{ \chi\sp{2}\sb{\rm red}}
                    \else\math{\chi\sp{2}\sb{\rm red}}\fi}
\newcommand\Teq{\ifmmode{T\sb{\rm eq}}\else$T$\sb{eq}\fi}
\newcommand\mjup{\ifmmode{M\sb{\rm Jup}}\else$M$\sb{Jup}\fi}
\newcommand\rjup{\ifmmode{R\sb{\rm Jup}}\else$R$\sb{Jup}\fi}
\newcommand\msun{\ifmmode{M\sb{\odot}}\else$M\sb{\odot}$\fi}
\newcommand\rsun{\ifmmode{R\sb{\odot}}\else$R\sb{\odot}$\fi}
\newcommand\mearth{\ifmmode{M\sb{\oplus}}\else$M\sb{\oplus}$\fi}
\newcommand\rearth{\ifmmode{R\sb{\oplus}}\else$R\sb{\oplus}$\fi}

\shorttitle{ZTF Image Processing}
\shortauthors{Laher {\em et al.}}

\begin{document}

\title{Processing Images from the Zwicky Transient Facility}

\author{Russ R. Laher\altaffilmark{1},
Frank J. Masci\altaffilmark{1},
Steve Groom\altaffilmark{1},
Benjamin Rusholme\altaffilmark{1},
David L. Shupe\altaffilmark{1},
Ed Jackson\altaffilmark{1},
Jason Surace\altaffilmark{1},
Dave Flynn\altaffilmark{1},
Walter Landry\altaffilmark{1},
Scott Terek\altaffilmark{1},
George Helou\altaffilmark{1},
Ron Beck\altaffilmark{1},
Eugean Hacopians\altaffilmark{2},\\
Umaa Rebbapragada\altaffilmark{3},
Brian Bue\altaffilmark{3},
Roger M. Smith\altaffilmark{4},
Richard G. Dekany\altaffilmark{4},
Adam A. Miller\altaffilmark{5},
S. B. Cenko\altaffilmark{6},\\
Eric Bellm\altaffilmark{7},
Maria Patterson\altaffilmark{7},
Thomas Kupfer\altaffilmark{8},
Lin Yan\altaffilmark{8},
Tom Barlow\altaffilmark{8},
Matthew Graham\altaffilmark{8},
Mansi M. Kasliwal\altaffilmark{8},
Thomas A. Prince\altaffilmark{8},
and
Shrinivas R. Kulkarni\altaffilmark{8}
}

\affil{\sp{1} IPAC, Mail Code 100-22, Caltech, 1200 E. California Blvd., Pasadena, CA 91125, U.S.A.\\
         \sp{2} Anre Technologies Inc., 3115 Foothill Blvd., Suite M202, La Crescenta, CA 91214, U.S.A.\\
         \sp{3} Jet Propulsion Laboratory, California Institute of Technology, Pasadena, CA 91109, U.S.A.\\
         \sp{4} Caltech Optical Observatories, California Institute of Technology, Pasadena, CA 91125, U.S.A.\\
         \sp{5} Center for Interdisciplinary Exploration and Research in Astrophysics, Northwestern University, Evanston, IL 60208, U.S.A.\\
         \sp{6} Astrophysics Science Division, NASA Goddard Space Flight Center, Code 661, Greenbelt, MD 20771, U.S.A.\\
         \sp{7} Department of Astronomy, University of Washington, Seattle, WA 98195, U.S.A.\\
         \sp{8} Division of Physics, Mathematics, and Astronomy, California Institute of Technology, Pasadena, CA 91125, U.S.A.}

\email{laher@ipac.caltech.edu}

\begin{abstract}
  The Zwicky Transient Facility is a new robotic-observing
  program, in which a newly engineered 600-MP digital camera with a pioneeringly large field of view,
  47~square degrees, will be installed into the 48-inch Samuel Oschin
  Telescope at the Palomar Observatory.  The camera will generate
  $\sim 1$~petabyte of raw image data over three years of operations.  
  In parallel related work, new hardware and
  software systems are being developed to process these data in real
  time and build a long-term archive for the processed products.  
  The first public release of archived products is
  planned for early 2019, which will include processed images and astronomical-source catalogs
  of the northern sky in the $g$ and $r$ bands.  Source catalogs based on two different methods will be 
  generated for the archive: aperture photometry and point-spread-function fitting.
\end{abstract}


\section{INTRODUCTION}
\label{introduction}

The Zwicky Transient Facility (ZTF\footnote{http://www.ptf.caltech.edu/ztf}) is a new program for ground-based,
optical, time-domain astronomy \citep{2015AAS...22532804B}, which goes well beyond its 
immensely successful predecessor programs, the Palomar
Transient Factory (PTF\footnote{http://www.ptf.caltech.edu}) and eponymous intermediate program (iPTF).
A digital camera specially developed for the ZTF, with a faster
readout time and a much larger field of view, along with other
enhancements relative to the PTF camera, such as a fast 
exposure shutter, is nearing the completion of
its development phase, and installation into the 48-inch Samuel Oschin
Telescope at the Palomar Observatory is planned for late summer of 2017.  
Substantial telescope and dome upgrades are part of this effort,
including faster drives and improved optics.  
In addition, the parallel development of hardware and software
systems that can handle real-time post-processing of the massive amount of image data from the camera in
normal operations, as well as the development of a long-term archive for the
processed products,
is underway.  The data processing will be performed at IPAC,
and the ZTF archive will be hosted by the NASA/IPAC Infrared Science Archive
(IRSA\footnote{http://irsa.ipac.caltech.edu}).  The data-processing system and archive will rely heavily on relational
databases to consolidate and warehouse information that can be later 
queried in a historical context for the discovery of astrophysical
transient phenomena.  The major science goals for the ZTF include discovering young supernovae,
searching for electromagnetic counterparts to gravitational wave
sources, identifying stellar variables, and detecting near-Earth
asteroids. 

In Section~\ref{sec:project}, we give background information on the ZTF project, 
some camera details, and highlights of the observing strategy.
In Section \ref{sec:processing}, the focus of this paper, we outline our design
of the ZTF data-processing system and discuss our recent performance-test results. 
We emphasize that the system design is still evolving and the results up to
this point in time are preliminary.  We briefly describe the ZTF
archive in Section~\ref{sec:archive}.  Finally, we provide a summary
and outlook in Section~\ref{sec:conclusions}.

\section{PROJECT BACKGROUND}
\label{sec:project}

\subsection{Programmatics}
\label{sec:program}

The ZTF's precedessor programs, PTF and iPTF, have led to the publication of 185 papers in
peer-reviewed journals (at the time of this writing -- more PTF/iPTF
papers are forthcoming).\footnote{The published papers are
  tracked at http://www.ptf.caltech.edu/iptf}  
As an example, a paper on a regular type II supernova based in part on PTF
data was recently published
in {\it Nature Physics} \citep{YaronEtAl2017SN}.
The iPTF project
officially ended in March of 2017.  A separate enterprise has been underway since
then to reprocess the PTF Galactic-plane data for inclusion in a
future public data release.

The ZTF is a consortium of several U.S. and global institutions, known as
the ``collaboration''.\footnote{http://www.ptf.caltech.edu/ztf}  
Both private funds from the members and public (National Science
Foundation) have been received for this project.  Work on the ZTF began in
earnest in 2014, and the current plan is for the ZTF to collect data for
three years after regular operations begin.

Data from the ZTF will fuel research projects for students, postdocs,
and scientists at institutions in the collaboration and the public at large.
There is also a related new initiative for undergraduate research known as the
Summer Undergraduate Astronomy Institute \citep{2017AAS...22931402P}.

\subsection{Camera}
\label{sec:camera}

Detailed engineering papers on the development of the ZTF camera are
given by \citet{SmithEtAl2014ZTF} and \citet{2016SPIE.9908E..5MD}.
A definitive instrument paper is in preparation by Dekany et
al., and is due out by end of 2017 or early 2018.
First light with the camera permanently mounted to the telescope is
scheduled for October 2017.

\begin{figure*}
\centering
\includegraphics[scale=0.38]{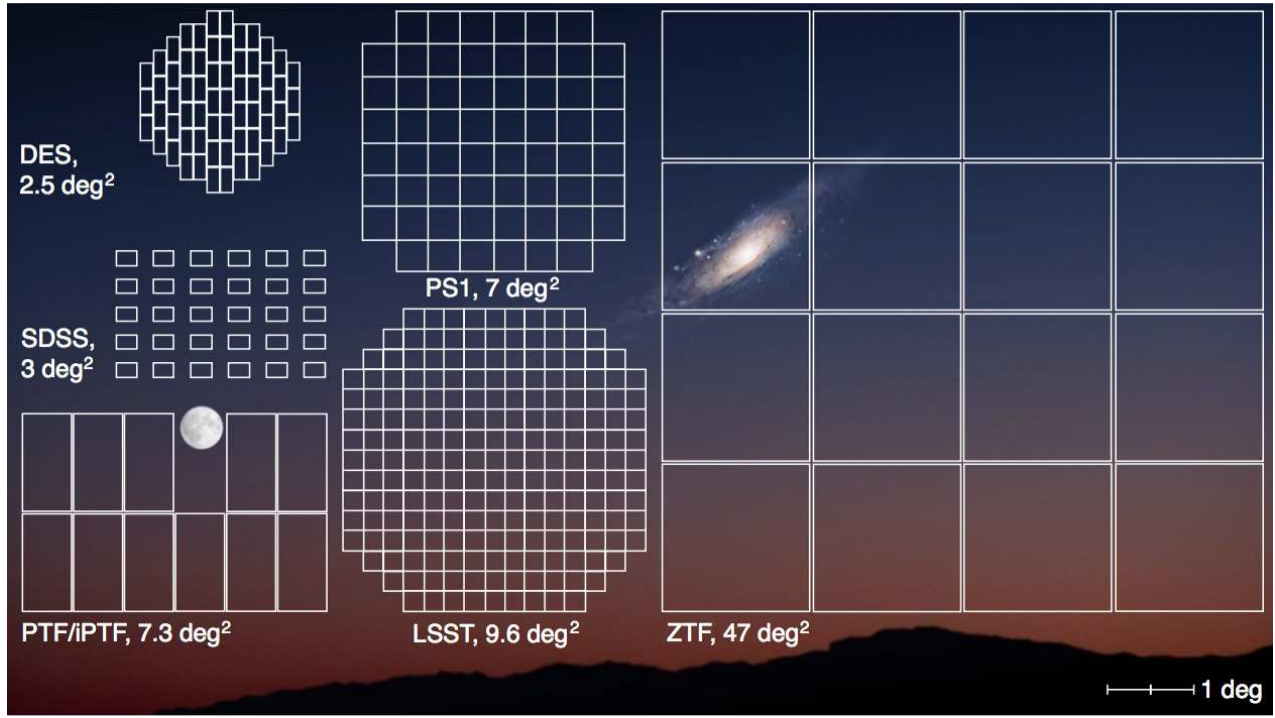}
\caption{Field of view of the ZTF camera compared to that of other
  large-survey cameras.  The Moon and the Andromeda Galaxy (Messier 31) are
  shown to scale.  (With the permission of Joel Johansson.) }
\label{fig:fov}
\end{figure*}

The ZTF-camera field of view is truly groundbreaking, and it will enable
imaging of the entire Palomar sky each night.
Figure~\ref{fig:fov} compares it to the field of view of other 
large-survey cameras, either currently in operation or under development.
Table~\ref{table:attributes} lists the salient attributes of the
camera for the planned observing.

\begin{table}[ht]
\centering
\caption{\label{table:attributes} ZTF camera and nominal observing attributes}
\begin{tabular}{ll}
\hline
\hline
Attribute                           & Value           \\
\hline
Field of view & 47 square degrees\\
Pixel scale & 1 arc second per pixel\\
Pixel size & 15 $\mu$m\\
CCD readout channels & 4\\
Exposure time & 30 seconds\\
Readout time & $\approx 10$ seconds\\
Slew \& settle time & $+5$ seconds\\
Optical filters & ZTF $g$, $r$, and $i$ \\
Limiting magnitude & 20.4 ($r$-band, 5 $\sigma$)\\
Mosaic of CCDs & $4 \times 4$ layout (16 CCDs)\\
Pixels per CCD & $\sim 6$K $\times$ 6K\\
Total number of pixels & $\approx 600$ megapixels\\
\hline
\end{tabular}
\end{table}

The median image quality delivered by the 48-inch telescope optics on
Palomar mountain is $\approx 2$~arc~seconds.  This
motivated the design choice of 1 arc second Nyquist sampling for the camera.

\subsection{Observing Strategy}
\label{sec:strategy}

The acquisition of science exposures will occur throughout each observing
night, weather permitting.
We are budgeting for 260 good-weather nights in our estimates.  Depending 
on the time of year, we expect to acquire 600-800 science exposures 
per night, assuming one exposure taken every 45 seconds, which 
allows for 30-second camera exposures and the fixed readout time 
and concurrent slew \& settle times of the camera and telescope (see Table~\ref{table:attributes}). 
Scanning at least 3760~square degrees of the sky per hour will be possible.

Calibration exposures (biases, dome flats, darks, etc.) will be acquired
during the day.  Fresh calibration products will be made before
the nightly processing of science exposures begins.

During the first year and a half of operations, ZTF will conduct two general-purpose public surveys:  a three-night cadence survey of the visible Northern Sky, and a nightly sweep of the Galactic Plane.  For both programs, fields will be visited twice each night they are observed, with approximately one hour separation between a $g$-band exposure and an $r$-band exposure.

\section{DATA PROCESSING}
\label{sec:processing}

\citet{LaherEtAl2014PTFProc} describe the image-processing system for
the ZTF's precedessor programs, PTF and iPTF.  This experience is leveraged in our
new design for the ZTF.  Below we give an overview of the ZTF Data
System.  A more detailed description will appear in a future publication.

The science exposures will be processed in real time, throughout the
observing night.  This requires a data-processing system that can
keep up with the incoming data.

\begin{figure*}
\centering
\includegraphics[scale=0.41]{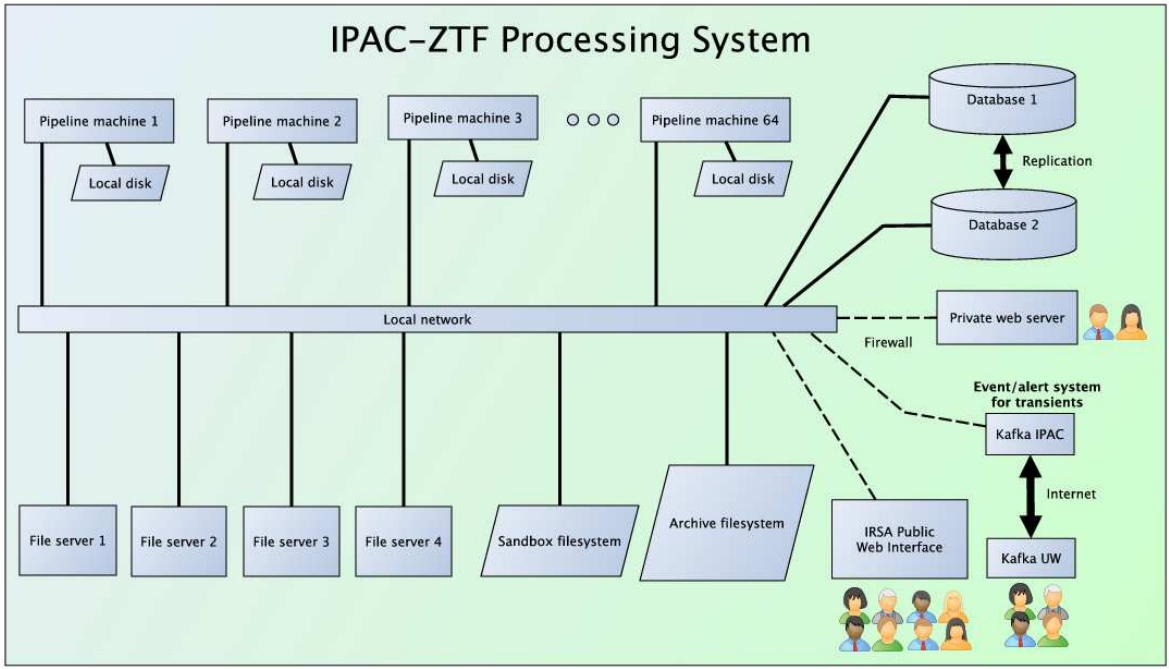}
\caption{ZTF data-processing system.}
\label{fig:processing}
\end{figure*}

Figure~\ref{fig:processing} depicts the ZTF
processing system that is currently under development at IPAC.  It features 64
pipeline machines, four file servers, two database machines (primary and
secondary/backup), a private web server, a Kafka\footnote{https://kafka.apache.org/documentation}
cluster, and an IRSA public web
interface. The local network is proficient at data rates of up to 10 gigabits per second.

The pipeline machines are for running ZTF real-time pipelines in
parallel.  This part of the system is inherently scalable, and more machines can
be added as needed.  The CPU of each pipeline machine has 16 cores (2
threads per core), and 16 pipeline instances per machine are typically
run.  The open-source, workload-manager software, SLURM\footnote{https://slurm.schedmd.com}, is used to 
farm out pipeline instances to pipeline machines in the proper order
and number.
The real-time pipeline generates many interim
files at its various steps, and these files typically become inputs
to a subsequent step.  The interim files are written to the local
disks of pipeline machines for speed and avoidance of unnecessary
congestion on the local network.

The four file servers allow parallel file transfer across the local network.
Files copied from pipeline-machine local disks to and from the sandbox
filesystem are expected to generate the most traffic.  Products
ultimately copied to the archive filesystem will add marginally to the load.  
This part of the system is also scalable and more machines could
potentially be added to service additional load.

Our data-processing system features a
PostgreSQL\footnote{https://www.postgresql.org} 
relational database running on the
primary database machine, equipped with 384 gigabytes of memory.
Since this element is not scalable, careful, disciplined database
design, and both hardware and software tuning have proven to be absolutely
essential.  Database tables and indexes are strategically micromanaged
so that heavily accessed data are stored on solid-state devices (SSDs)
as opposed to spinning disks.  The large database-machine memory and
judicious database tuning ensure a very
high cache-hit rate.
Daily database vacuuming is required for regular maintenance.
The database is replicated onto a second database machine, so that
design-team members can query it without affecting the primary
database.  Candidate transients are stored in database tables 
partitioned by observing date and readout channel.  A set-up process
that creates database views over the correct candidate partitions is executed daily.

The web server will allow ZTF personnel to perform quality analysis on the raw data and processed products.  It will also serve as a staging ground for initial access for archive products as well as disseminate products for solar system science.

The Kafka cluster is a set of machines at IPAC and a set of machines
at the University of Washington (UW) that facilitates the distribution of
transient-event alerts to the ZTF collaboration and eventually the
community.  Once the event data has been sent to Kafka,
which is facilitated by pipeline packaging the data in Avro
format,\footnote{E.g., https://avro.apache.org/docs/1.8.2/} 
the Kafka software handles mirroring between IPAC and UW over the Internet.

\begin{figure}
\centering
\includegraphics[scale=0.42]{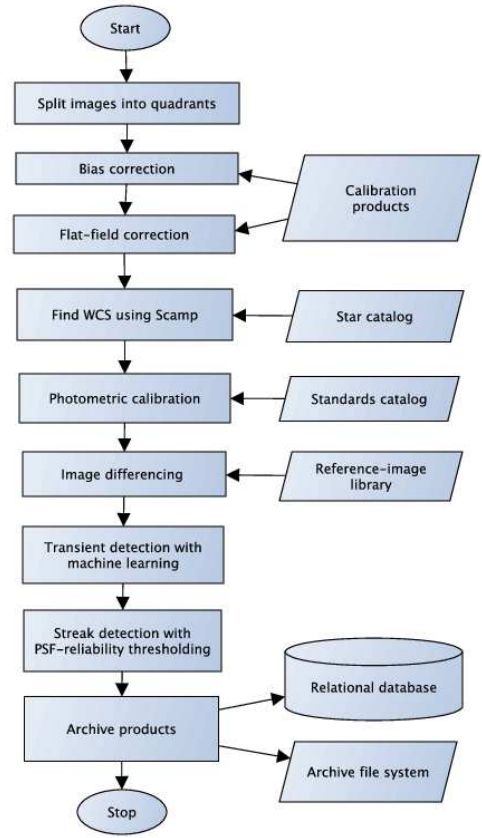}
\caption{Flowchart of the ZTF real-time pipeline.}
\label{fig:pipe}
\end{figure}

Figure~\ref{fig:pipe} gives a flowchart of the ZTF real-time pipeline.
The CCD-image files received from Palomar consist of one multi-extension FITS
file for each CCD and exposure.  These are split into readout-channel
quadrants for smaller images to facilitate the subsequent parallel
processing, in which the quadrant images are processed independently.  
Bias and flat-field corrections are applied to the applicable raw
images.
The processing involves instrumental calibration, which consists of
astrometric and photometric calibration, followed by image
differencing with a reference image for finding transients.
Scamp\footnote{https://www.astromatic.net/software/scamp} is used in
conjuction with the GAIA catalog \citep{GaiaDR1} to perform the astrometry and compute
a World Coordinate System for each science image.  Absolute photometry is done using the
Pan-STARRS DR1 catalog as the standard \citep{2017AAS...22923707F}.
Reference images, which are needed for
image differencing, are made within the first few months of operation.
The image differencing includes a combination of the steps described in \citet{2017PASP..129a4002M}, as well as an implementation of the ZOGY algorithm \citep{ZOGY2016}.  
Candidate transient events are detected in
thresholded difference images, and then vetted with machine learning
provided by JPL, for example, using the real/bogus framework of \citet{2012PASP..124.1175B}.  
Asteroid streaks are found in the difference images
using the {\it findstreaks} module \citep{2017PASP..129c4402W}, and
subsequently checked for reliability using a new deterministic 
algorithm that accounts for the point-spread function of
the streak.

\begin{figure*}
\centering
\includegraphics[scale=0.5]{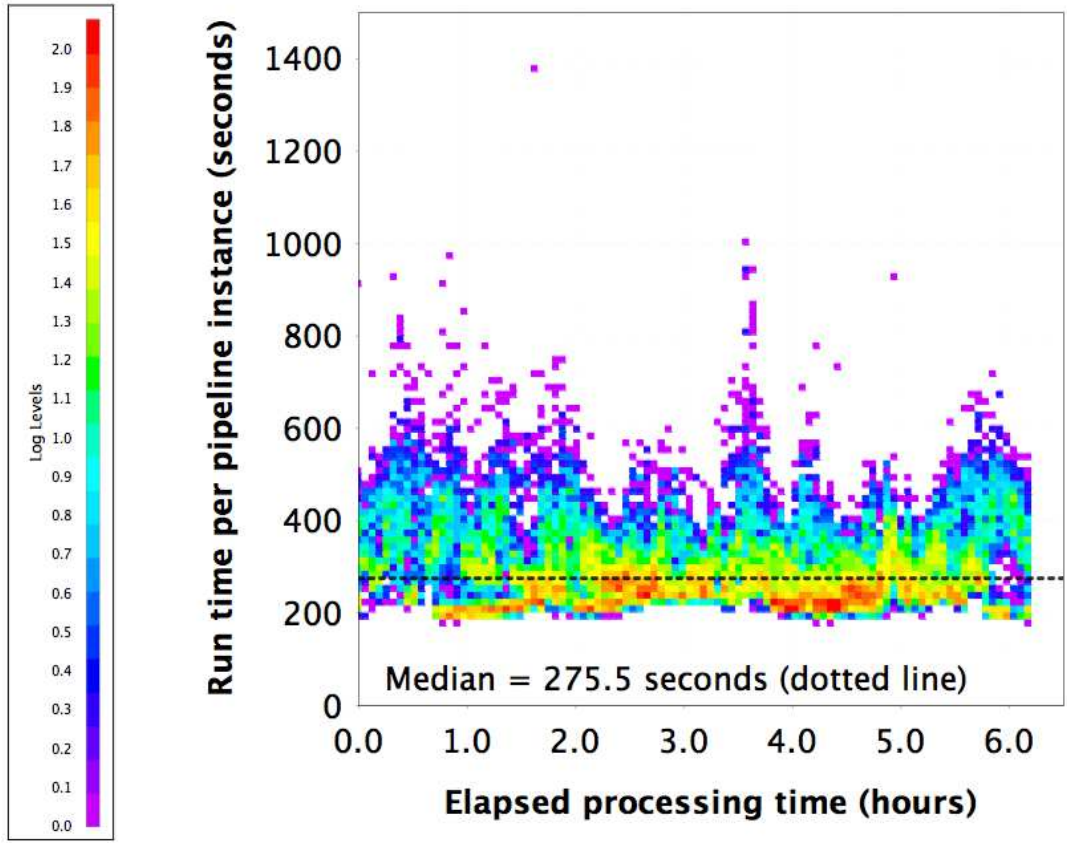}
\caption{Preliminary ZTF benchmark-test results.}
\label{fig:test}
\end{figure*}

Figure~\ref{fig:test} gives preliminary performance test results for the ZTF
real-time pipelines processing a night's worth of simulated data.
The density plot includes 28,505~independent pipeline instances
running on 32 pipeline machines, and shows that a
night spanning $\sim 7.5$~hours can be processed in $\sim 6.2$~hours,
which is better than real time.  The median run time for a pipeline
instance in this test is 275.5~seconds.  The primary factors that affect the
pipeline run times include: 1)~The number of astronomical sources
extracted from the images (e.g., Galactic-plane observations are
much more stressing); and 2)~The fraction of images for the night that
actually have reference images available (this was $\sim 75$\% for our
simulated data).  The simulated data for the test includes random
transients at a rate of 10 per CCD.  Future tests will involve simulated
transient rates that are up to $8\times$ higher.

\section{PRODUCT ARCHIVE}
\label{sec:archive}

As illustrated in Figure~\ref{fig:processing},
IRSA
will set up infrastructure for a long-term archive and a web interface 
for the distribution of ZTF-survey products
to the collaboration and public.
The web interface will have similar functionality as those available to IRSA archives
for other projects.  The entire archive is expected to grow to 
$\sim3$~petabytes by the end of ZTF's three-year program.
Table~\ref{table:products} lists the product types that will be archived,
which will include both images and astronomical-source catalogs.  Two different methods will be 
used to generate source catalogs for the archive, namely, aperture
photometry and point-spread-function (PSF) fitting.

The first public release of archived products is planned for early 2019.
This release will include only products from the ZTF public surveys.  
Products from collaboration observing programs will be included in later releases.

\section{CONCLUSIONS}
\label{sec:conclusions}

This paper gives an overview of the ZTF project, camera, observing
strategy, data-processing system, preliminary benchmark results,
and product archive, with an
emphasis on describing the data-processing system and archive.
In the coming months, the data-processing system will be exercised 
on real data from the observatory-mounted camera.  Building upon our past
successes gives us confidence in a favorable outcome for this endeavor.  
Exciting new discoveries await astronomers analyzing ZTF products!

\begin{table}[ht]
\centering
\caption{\label{table:products} ZTF archive products}
\begin{tabular}{ll}
\hline
\hline
Product\\
\hline
Raw images\\
Processed images\\
Image masks\\
Difference images\\
SExtractor catalogs\\
PSF-fit catalogs (DAOPHOT)\\
Reference images, catalogs, etc.\\
Calibration products\\
Light curves\\
\hline
\end{tabular}
\end{table}

\acknowledgments
We are grateful to Joel Johansson of the Oskar Klein Center at Stockholm University and the Weizmann Institute of Science for permission to use the original creative work shown in Figure~\ref{fig:fov}.

The ZTF is supported by a collaboration including Caltech, IPAC, the
Weizmann Institute of Science, the Oskar Klein Center at Stockholm
University, the University of Maryland, Deutsches
Elektronen-Synchrotron and Humboldt University, Los Alamos National
Laboratories, the TANGO Consortium of Taiwan, the University of
Wisconsin at Milwaukee, Lawrence Berkeley National Laboratories,
and the University of Washington.

Matching support from the National Science Foundation MSIP program will enable public ZTF surveys, data releases, and annual summer schools.

\end{document}